\documentclass[a4paper,onecolumn,showpacs]{revtex4}
\usepackage{amsmath}
\usepackage{graphicx}
\usepackage[english]{babel} 
\usepackage[latin1]{inputenc} 
\usepackage[T1]{fontenc}

\begin{document}
\title{Study of universality crossover in the contact process}
\author{Wellington G. \surname{Dantas}}
\email{wgd@if.uff.br}  
\author{J\"urgen F. \surname{Stilck}}
\email{jstilck@if.uff.br}  
\affiliation{Instituto de F\'{\i}sica, Universidade Federal 
Fluminense, \\
Campus da Praia Vermelha,  \\
Niter\'oi, RJ, 24.210-340, Brazil.}

\date{\today}


\begin{abstract}
We consider a generalization of the contact process stochastic model,
including an additional autocatalitic process. The phase diagram of this model
in the proper two-parameter space displays a line of transitions between an
active and an absorbing phase which starts at the critical point of the
contact process and ends at the transition point of the voter model. Thus, a
crossover between the directed percolation and the compact percolation
universality classes is observed at this latter point. We study this crossover
by a variety of techniques. Using supercritical series expansions analyzed
with partial differential approximants, we obtain precise estimates of the
crossover behavior of the model. In particular, we find an estimate for the
crossover exponent $\phi=2.00 \pm 0.02$. We also show arguments that support
the conjecture $\phi=2$.
\end{abstract}

\maketitle

\section{Introduction}
\label{intro}
The phase transitions exhibited by stochastic models with absorbing states
have attracted much attention in recent years, particularly in order to
identify and understand the aspects which determine the universality classes
in those models. Most of these models have not been solved exactly, but a
variety of approximations allow quite conclusive results regarding their
critical properties. Stochastic models are, of course, well fitted for
simulations, but closed form approximations and other analytical approaches
have also been useful in investigating their behavior \cite{md99}.

One of the simplest and most studied model of this type is the contact process
(CP), which was conceived as a simple model for the spreading of an
epidemic and proven to display a continuous transition between the absorbing
and an active state, even in one dimension \cite{h74}. Actually, it was found
that the CP is equivalent to other models such as Schl\"ogl's lattice model
for autocatalytic chemical reactions \cite{s72} and Reggeon Field Theory (RFT)
\cite{gt}. The CP belongs to the direct percolation (DP) universality class,
together with others models such as the Ziff-Gulari-Barshad model of catalysis
\cite{zgb86} and branching and annihilating walks with an odd offspring
\cite{tty92}. The {\em DP conjecture} states that all phase transitions between
an active and an absorbing state in models with a scalar order parameter,
short range interactions and no conservation laws belong to this class
\cite{j81}. This conjecture was verified in all cases studied so far
\cite{h00}.

Here we study a generalization of the CP, with an additional parameter, so
that the CP transition point becomes a critical line. Since the symmetry
properties of this generalized model are the same of the CP, it is expected
that this critical line should belong to the DP universality class. However,
at one point of this line the model is equivalent to the zero temperature
Glauber model \cite{g63}, also called the voter model \cite{l85}, which
displays a spin 
inversion (or particle-hole) symmetry and therefore belongs to another
universality class. Thus the critical line in the phase diagram of the
generalized model starts at the CP model and ends at the voter model, a
crossover between the two universality classes being observed. The voter model
belongs to the compact percolation universality class, and also corresponds to
a limiting point in the phase diagram of the Domany-Kinzel cellular automaton,
where an exact solution is possible \cite{dk84}. Thus, the exact critical
exponents are known for this model. In a study of models with several
absorbing states \cite{h97} simulational results are shown for the model we
are considering here and a study of the crossover between direct percolation
and compact direct percolation may be found in \cite{mdh96}, motivated by
the possibility to explain the non-universality in models with several
absorbing states as a surface effect. Also, the shape of the critical line
close to the CDP endpoint in the Domany-Kinzel automaton was studied in detail
\cite{kt95,tikt95,ikb01}, and these results are compared to our findings in
the conclusion.  Some physical motivation for the model
we are studying here 
might be given. In the contact process, the additional term might be
understood as an enhancement of the possibility of a sick individual to
recover proportional to the number of its first neighbor which are
healthy. However, our motivation to study the model is centered on its
simplicity and the universality class crossover present in its phase diagram. 

In section \ref{model} we define the model and explain how supercritical
series expansions may be obtained for it. The coefficients of the two-variable
series for the survival probability up to order 25 are given. Section \ref{res}
contains the description and the results of the Pad\'e and PDA estimates for
the model, with emphasis on the multicritical behavior in the voter model
limit. Final discussions and the conclusion may be found in section
\ref{conclu}.  

\section{Definition of the model and calculation of the coefficients of the
supercritical series for the survival probability}
\label{model}
The model is defined on a one-dimensional lattice with $N$ sites and periodic
boundary conditions. Each site is occupied either by a particle A or a
particle B, no holes are allowed. The microscopic state of the model may thus
be described by the set of binary variables
$\eta=(\eta_1,\eta_2,\ldots,\eta_N)$, where $\eta_i=0$ or $1$ if site $i$ is
occupied by particles B or A, respectively. 

The model evolves in time according to the following Markovian rules:
\begin{enumerate}
\item A site $i$ of the lattice is chosen at random.
\item If the site is occupied by a particle B, it becomes occupied by a
particle A with a transition rate equal to $p_a n_A/2$, where $n_A$ is the
number of A particles in the sites which are first neighbors to site $i$.
\item If site $i$ is occupied by a particle A, it may become occupied by a
particle B through two processes:
\begin{itemize}
\item Spontaneously, with a transition rate $p_c$.
\item Through an autocatalytic reaction, with a rate $p_bn_B/2$, where $n_B$
is the number of B particles in the sites which are first neighbors to site
$i$. 
\end{itemize}
\end{enumerate}

We define the time in such a way the the non-negative parameters $p_a$, $p_b$,
and $p_c$ obey the normalization $p_a+p_b+p_c=1$. We may then discuss the
behavior of the model in the $(p_a,p_c)$ plane without loss of generality.

The probability $P(\eta,t)$ to find the system in state $\eta$ at 
time $t$ obeys the master equation
\begin{eqnarray}
\label{eq1}
\frac{\partial P(\eta,t)}{\partial t} = 
\sum_{i=1}^N [w_i(\eta^i)P(\eta^i,t)
-w_i(\eta)P(\eta,t)]
\label{me}
\end{eqnarray}
where $\eta^i$ corresponds to the following configuration
\begin{eqnarray}
\label{eq2}
\eta^i\equiv(\eta_1,...,1-\eta_i,...,\eta_N)
\end{eqnarray}
and $w_i(\eta)$ is the transition rate of the model, given by
\begin{eqnarray}
\label{eq3}
w_{i}(\eta) = 
\frac{\mu}{2}(1-\gamma\eta_i)\sum_{\delta}
\eta_{i+\delta}+\eta_i,
\end{eqnarray}
where $\mu = p_a/(1-p_a)$, $\gamma = (1-p_c)/p_a$, and the sum is 
over first neighbors of site $i$.  

It may be useful to remark that this model may be mapped to a spin 
system if we describe sites occupied by A and B particles by an Ising 
spin variables $\sigma_i=1$ and $\sigma_i=-1$, respectively. 
In these variables, 
the transition rate will be given by
\begin{eqnarray}
\label{eq6}
w_{i}(\sigma) = 
\frac{\alpha}{2}\left[1+\beta \sigma_i-\frac{1}{2}(\epsilon 
\sigma_i+\xi)
\sum_{\delta}\sigma_{i+\delta}\right],
\end{eqnarray}
where $\alpha =(p_a+p_b+2p_c)/2,\beta=(p_a-p_b-2p_c)/(pa+p_b+2p_c),
\epsilon=(p_a+p_b)/(p_a+p_b+2p_c)$, and  
$\xi=(p_a-p_b)/(p_a+p_b+2p_c)$.

In two particular cases, this model corresponds to well known models. If we
make $p_b=0$ or $\gamma=1$ contact  
process is recovered \cite{h74}:
\begin{eqnarray}
\label{eq7}
w_{i}^{(CP)}(\eta)=\frac{\mu}{2}(1-\eta_i)\sum_{\delta}
\eta_{i+\delta}+\eta_i.
\end{eqnarray}
If now we take $p_a=p_b$ and $p_c=0$ in the spin formulation of the 
model, the zero temperature linear Glauber model \cite{o03}, also 
known as the voter model, is recovered \cite{l85}:
\begin{eqnarray}
w_{i}^{(LGM)}(\sigma) = \frac{\alpha}{2}\left[1-\frac{1}{2}\sigma_i
\sum_{\delta}\sigma_{i+\delta}\right].
\end{eqnarray}

\begin{center}
\begin{figure}[h!]
\includegraphics[scale=0.7]{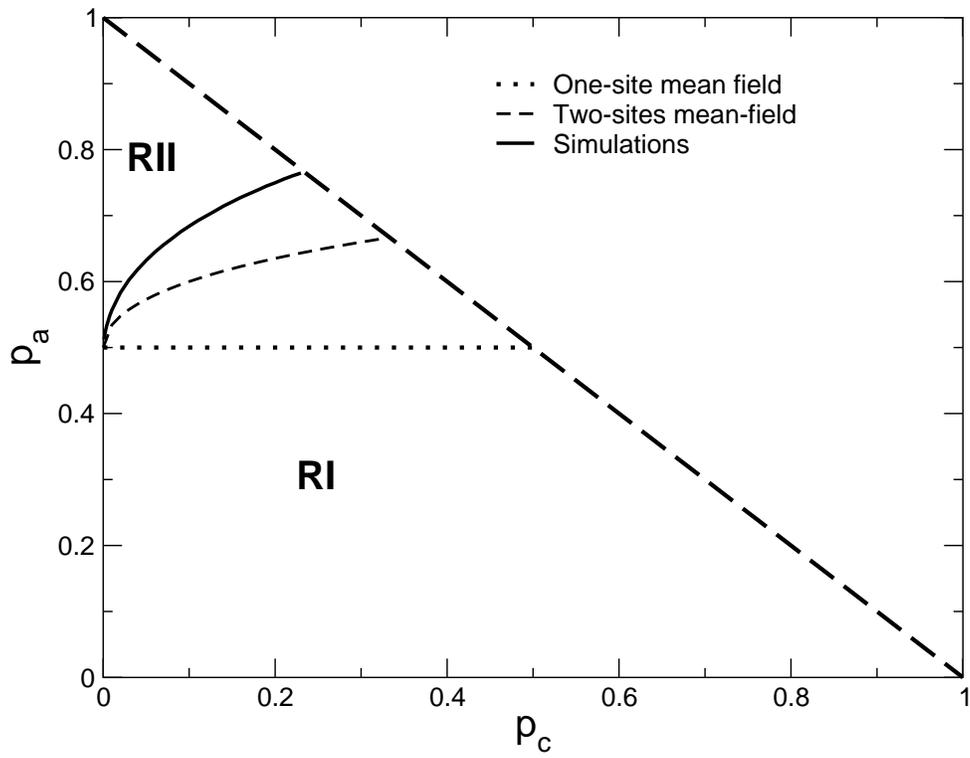}
\vspace{0.4cm}
\caption{Phase diagram of the model obtained from mean-field approximations
and simulations. The physical region is $p_a+p_c \leq 1$ and the line
$p_a+p_c=1$ corresponds to the contact process. In the region labeled RI the
absorbing phase (all particles of type B) is stable, while in the region RII
an active phase is stable, with a nonzero density of particles A.}
\label{f1}
\end{figure}
\end{center}

This model has been studied using mean-field approximations \cite{dts03}, as
well as simulations \cite{h97,dts03}. For $p_c>0$, the stationary state at low
values of $p_A$ 
corresponds to the absorbing state, where the density of A particles
$\rho_A=<N_A>/N$ vanishes. As $p_a$ is increased, a continuous phase
transition occurs and an active stationary state ($\rho_A>0$) is stable at
high values of $p_a$. Thus a critical line is present in the phase diagram
starting at $(p_a=0.767325(6),p_c=1-p_a=0.232674(4))$ \cite{dj91} (contact
process), and ending at $(p_a=1/2,p_c=0)$ (linear Glauber model), where the
transition is discontinuous and between two absorbing states ($\rho_A=1$ and
$\rho_B=1$). In figure \ref{f1} results from mean-field calculations and
simulations for the phase diagram are shown \cite{dts03}. While it is expected
that the critical exponents at the whole critical line are the ones of the
directed percolation (DP) universality class, a crossover to the compact
directed percolation (CDP) universality class exponents happens as $p_c$
vanishes \cite{h00}. Thus, we may recognize the point $(p_a=1/2,p_c=0)$ of the
phase diagram as a multicritical point. Using supercritical series expansions,
for the survival probability, we will study the multicritical singularity in
the neighborhood of this point.

Now let us develop a two-variable supercritical series expansion for the
model. We follow closely the operator formalism presented in the paper by
Jensen and Dickman on series for the CP process and related models
\cite{dj91}. We may then represent the microscopic configurations of the
lattice by the direct product of kets
\begin{equation}
| \eta \rangle= \bigotimes_i | \eta_i \rangle,
\end{equation}
which are defined to be orthonormal
\begin{equation}
\langle \eta | \eta^\prime \rangle = \prod_i \delta_{\eta_i,\eta_i\prime}.
\end{equation}
Now we may define A particles creation and annihilation operators for the site
$i$: 
\begin{eqnarray}
A_i^\dagger |\eta_i \rangle &=&(1-\eta_i)|\eta_i+1\rangle,\nonumber \\
A_i |\eta_i\rangle &=& \eta_i|\eta_i-1\rangle.
\end{eqnarray}
In this formalism, the state of the system at time $t$ is
\begin{equation}
|\psi(t) \rangle = \sum_{\{\eta\}} p(\eta,t) |\eta \rangle.
\end{equation}
If we define the projection onto all possible states as
\begin{equation}
\langle \; | \equiv \sum_{\{\eta\}} \langle\eta|,
\end{equation}
the normalization of the state of the system may be expressed as $\langle \;
|\psi \rangle =1$. In this notation, the master equation for the evolution of
the state of the system (Eq. \ref{me}) is
\begin{equation}
\frac{d |\psi(t)\rangle}{d t}=S|\psi(t)\rangle.
\label{me1}
\end{equation}
The evolution operator $S$ may be expressed in terms of the creation and
annihilation operators as $S=\lambda S_0 + V$ where
\begin{eqnarray}
S_0 &=& \sum_i [\alpha(2-A_{i-1}^\dagger A_{i-1} +
A_{i+1}^\dagger A_{i+1})+1](A_i-A_i^\dagger A_i), \\
V &=& \sum_i (A_i^\dagger + A_i^\dagger A_i -1)(A_{i-1}^\dagger A_{i-1} +
A_{i+1}^\dagger A_{i+1}),
\end{eqnarray}
and the new parameters 
\[
\lambda \equiv \frac{2p_c}{p_a}
\]
and
\[
\alpha \equiv \frac{p_b}{2p_c}
\]
were introduced.

We notice that the operator $S_0$ only annihilates A particles (transitions $A
\to B$), while the operator $V$ acts in the opposite way, generating
transitions $B \to A$. Thus, for small values of the parameter $\lambda$ the
creation of A particles is favored, and the decomposition above is convenient
for a supercritical perturbation expansion. Let us show explicitly the effect
of each operator on a configuration $(\mathcal{C})$. 
\begin{equation}
S_0({\mathcal{C}})=\alpha \sum_i
({\mathcal{C}}^\prime_i)+2\alpha \sum_j
({\mathcal{C}}^\prime_j) + \sum_k
({\mathcal{C}}^\prime_k)- [\alpha(r_1+2r_2)+r] ({\mathcal{C}}),
\label{as0}
\end{equation}
where the first sum is over the $r_1$ sites with A particles and one B
neighbor, 
the second sum is over the $r_2$ sites with A particles and two B neighbors
and the third 
sum is over all $r$ sites with A particles of the configuration
$(\mathcal{C})$. Configuration $({\mathcal{C}}_i^{\prime})$ is obtained 
replacing
the A particle at site $i$ by a B particle. The action of operator $V$ is
\begin{equation}
V({\mathcal{C}})=\sum_i ({\mathcal{C}}^{\prime \prime}_i)+2 \sum_j
({\mathcal{C}}^{\prime \prime}_j) -(q_1+2q_2)({\mathcal{C}}),
\label{av}
\end{equation}
where the first sum is over the $q_1$ sites with B particles and one A
neighbor, 
the second sum is over the $q_2$ sites with B particles and two A
neighbors. Configuration $({\mathcal{C}}_i^{\prime \prime})$ is obtained
replacing the B particle at site $i$ in configuration $(\mathcal{C})$ by a A
particle.

To obtain a supercritical expansion for the ultimate survival probability of A
particles, we start by remembering that in order to access the long-time
behavior of a quantity, it is useful to consider its Laplace transform. For
instance, the Laplace transform of the state of the system is
\begin{equation}
|\tilde{\psi}(s) \rangle = \int_0^\infty e^{-st} |\psi(t)\rangle,
\label{lt}
\end{equation}
and inserting the formal solution $|\psi(t)\rangle =e^{St} |\psi(0)\rangle$ of
the master equation \ref{me1}, we find
\begin{equation}
|\tilde{\psi}(s) \rangle = (s-S)^{-1} |\psi(0)\rangle.
\label{tpsi}
\end{equation}
The stationary state $|\psi(\infty) \rangle \equiv \lim_{t \to \infty} |\psi(t)
\rangle$ may then be found noticing that 
\begin{equation}
|\psi(\infty) \rangle = \lim_{s \to 0} s |\tilde{\psi}(s) \rangle,
\end{equation}
which may be obtained integrating \ref{lt} by parts. A perturbative expansion
may be obtained assuming that $|\tilde{\psi}(s) \rangle$ may be expanded in
powers of $\lambda$ and using \ref{tpsi},
\begin{equation}
|\tilde{\psi}(s) \rangle = |\tilde{\psi}_0 \rangle+\lambda |\tilde{\psi}_1
\rangle +\lambda^2 |\tilde{\psi}_2 \rangle + \cdots = \frac{1}{s- V -\lambda
S_0 } |\psi(0) \rangle. 
\end{equation}
Since
\begin{equation}
\frac{1}{s- V -\lambda S_0 }= \frac{1}{s-V} \left[ 1 + \lambda \frac{1}{s-V}
S_0 
+ \lambda^2 \frac{1}{(s-V)^2} S_0^2 + \cdots \right],
\end{equation}
we arrive at
\begin{eqnarray}
|\tilde{\psi}_0 \rangle &=& \frac{1}{s-V} |\psi(0)\rangle \nonumber \\
|\tilde{\psi}_1 \rangle &=& \frac{1}{s-V} S_0 |\tilde{\psi}_0 \rangle
\label{rr} \\
|\tilde{\psi}_2 \rangle &=& \frac{1}{s-V} S_0 |\tilde{\psi}_1 \rangle
\nonumber \\
 &\vdots& 
\end{eqnarray}
The action of the operator $(s-V)^{-1}$ on an arbitrary configuration
$({\mathcal{C}})$ may be found noting that
\begin{equation}
(s-V)^{-1} ({\mathcal{C}})=s^{-1} ({\mathcal{C}})+ \frac{V}{s(s-V)}
({\mathcal{C}}), 
\end{equation}
and using the expression \ref{av} for the action of the operator $V$, we get
\begin{equation}
(s-V)^{-1} ({\mathcal{C}})= s_q \left\{ ({\mathcal{C}}) + (s-V)^{-1} \left[
\sum_i 
({\mathcal{C}}^{\prime \prime}_i)+2 \sum_j ({\mathcal{C}}^{\prime \prime}_j)
\right] \right\},
\label{sv}
\end{equation}
where the first sum is over the $q_1$ sites with B particles and one A
neighbor, the second sum is over the $q_2$ sites with B particles and two A
neighbors, and we define $s_q \equiv 1/(s+q)$, where $q=q_1+2 q_2$. 

It is convenient to adopt as the initial configuration a translational
invariant one with a single A particle (periodic boundary conditions are
adopted). Now we may notice in the recursive expression \ref{sv} that the
operator $(s-V)^{-1}$ acting on any configuration generates an infinite set of
configurations, and thus we are unable to calculate $| \tilde{\psi} \rangle$ in
a closed form. We may, however, calculate the extinction probability
$\tilde{p}(s)$, which corresponds to the coefficient of the vacuum state
$|0\rangle$. As happens also for the models related to the CP studied in
\cite{dj91} configurations with more than $j$ particles only contribute at
orders higher than $j$, and since we are interested in the ultimate survival
probability for A particles $P_\infty=1-\lim_{s \to 0} s \tilde{p}(s)$, $s_q$
may be replaced by $1/q$ in Eq. \ref{sv}. To illustrate the procedure, we will
perform the explicit calculation of the series for $\lim_{s \to
0}s\tilde{p}(s)$ up to third 
order in $\lambda$. We furthermore represent a configuration denoting by
$\bullet$ a site occupied by an A particle and by $\circ$ a site occupied by
a B particle. Thus $(\bullet \circ \bullet)$ denotes the translationally
invariant configuration $\sum_i A_i^\dagger A_{i+2}^\dagger |0 \rangle$. The B
particles situated to the left of the leftmore A particle  in a configuration
and the ones situated to the right 
of the rightmost A particle are omitted in the representation of this
configuration. The vacuum
state will be 
represented by $(0)$. As stated above, at $t=0$ the
system will be supposed to be in the configuration $|\psi(0) \rangle = \sum_i
A_i^\dagger |0 \rangle = (\bullet)$. Keeping configurations with up to three
particles, 
and omitting the global factor $1/s$, the first of the recursion relations
Eqs. \ref{rr} leads to 
\begin{equation}
|\tilde{\psi}_0 \rangle=\frac{1}{2}[(\bullet)+(\bullet
\bullet)+(\bullet\bullet\bullet)+ \cdots].
\end{equation}
The next step is the calculation of
$|\tilde{\phi}_0\rangle=S_0|\tilde{\psi}_0\rangle$. Now we need to keep only
configurations with up to two particle. The result is
\begin{equation}
|\tilde{\phi}_0\rangle=\frac{1}{2}[(2\alpha+1)(0)+(\bullet)+
\bullet\circ\bullet)+\cdots].
\end{equation}
Now we obtain $|\tilde{\psi}_1\rangle=(s-V)^{-1} |\tilde{\phi}_0\rangle$ for
$s=0$, resulting in
\begin{equation}
|\tilde{\psi}_1\rangle=\frac{1}{2}(2\alpha+1)(0)+\frac{1}{4}(\bullet)+
\frac{1}{4}(\bullet\bullet)+\frac{1}{8}(\bullet\circ\bullet)+\cdots.
\end{equation}
From this point on we will only show the results of each step, up to the
relevant numbers of A particles:
\begin{eqnarray}
|\tilde{\phi}_1\rangle &=& \frac{1}{4}(2\alpha+1)(0)+\frac{1}{4}
(2\alpha+2)(\bullet)+\cdots, \nonumber \\
|\tilde{\psi}_2\rangle &=& \frac{1}{4}(2\alpha+1)(0)+\frac{1}{8}
(2\alpha+2)(\bullet)+\cdots, \nonumber \\
|\tilde{\phi}_2\rangle &=& \frac{1}{8}(2\alpha+2)(2\alpha+1)(0)+\cdots,
\nonumber \\ 
|\tilde{\psi}_3\rangle &=&
\frac{1}{8}(2\alpha+2)(2\alpha+1)(0)+\cdots. \nonumber 
\end{eqnarray}
The first coefficients of the ultimate survival probability will then be given
by 
\begin{equation}
P_\infty=1-\frac{1}{2}(2\alpha+1)\lambda+-\frac{1}{4}(2\alpha+1)\lambda^2 +
\frac{1}{8}(2\alpha+2)(2\alpha+1) \lambda^3 + \cdots
\end{equation}

The algebraic operations above may be easily performed in a computer using a
proper algorithm. The
configurations are expressed as binary numbers and the coefficients as double
precision variables. Although we have tried to do the calculation representing
the coefficients as rational numbers, thus avoiding any roundoff errors since
all calculations were done with integers, we found that the denominators
increase very rapidly with the order of the calculations, and thus we were
unable to perform the calculations this way up to a reasonable order. With
rather modest computational resources (Athlon MP2200, double processor, 1Gb
memory) it is not difficult to calculate the coefficients up to order 25. The
required processing time amounts to about 6 hours, the limiting factor is
actually the memory required for the calculation. The maximum number of terms
(polynomials in $\alpha$) amounts to more than $44 \times 10^6$. We define the
coefficients $b_{i,j}$ as:
\begin{equation}
P_\infty=1-\frac{1}{2}(2\alpha+1) \lambda - \sum_{i=2}^{25}  \sum_{j=0}^{i-1}
b_{i,j} \lambda^i \alpha^j,
\end{equation}
and they are listed in Table \ref{coefs}. Up to order 24, our results are
numerically coincident with the supercritical series expansion for the
ultimate survival probability of the contact process \cite{dj91}, in the
particular case $\alpha=0$ (we remark that the variable $\lambda$ in the
supercritical expansion for the ultimate survival probability in reference
\cite{dj91} is half the variable $\lambda$ we use here). 

\section{Analysis of the series}
\label{res}
\begin{center}
\begin{figure}
\includegraphics[scale=0.7]{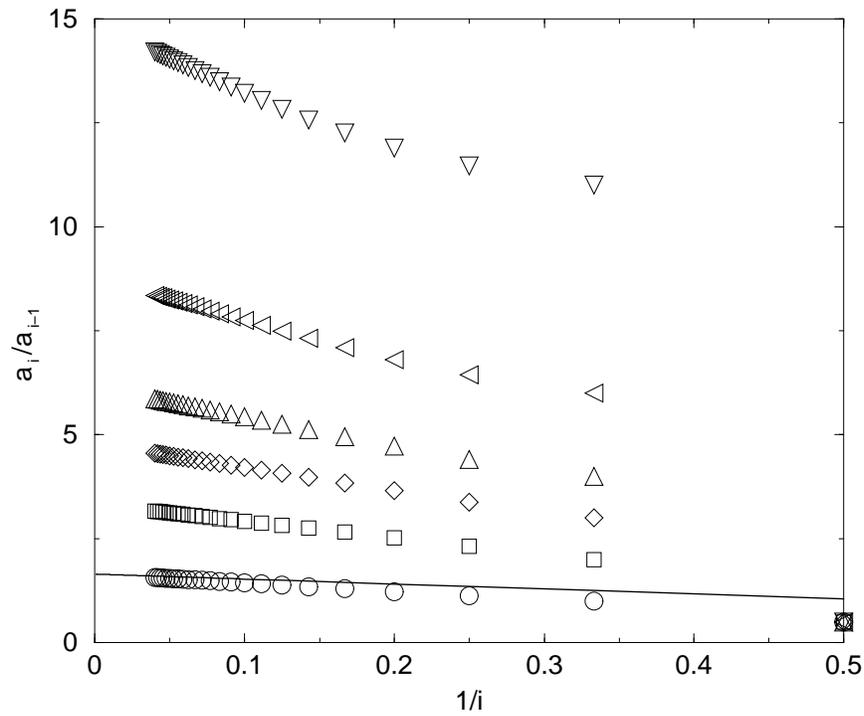}
\caption{Ratio of successive coefficients of the series expansion in $\lambda$
for fixed values of $\alpha$. The results shown are for $\alpha=0$ (circles),
$\alpha=1$, $\alpha=2$, $\alpha=3$ $\alpha=5$, and $\alpha=10$ (triangles
pointing down). The solid line shows the expected asymptotic behavior for
$\alpha=0$ (contact process).}
\label{rm}
\end{figure}
\end{center}

Let us consider initially the one variable series for fixed values of
$\alpha$
\begin{equation}
P_\infty=\sum_{i=0}^{25} a_i(\alpha) \lambda^i.
\end{equation}
As a preliminary test, we may apply the ratio method \cite{g89} to these
series. The 
results for $r_i=a_i/a_{i-1}$ as functions of $1/i$ are depicted in Figure
\ref{rm}. In the case $\alpha=0$ (circles), the asymptotic linear behavior
 $r_i \approx 1,6489(1-0.7231/i)$,
obtained from precise estimates for $\lambda_c$ and the exponent $\beta$ for
the contact process \cite{dj91}. In the figure it is apparent that the ratios
approach the asymptotic limit as $i$ is increased, as a matter of fact this
approach is close to linear in $1/i$. Thus, we may infer that for this model
the singularity which is closer to the origin is actually the physical
singularity. As the value of $\alpha$ is increased, one may notice that the
asymptotic linear behavior in $1/i$ for the ratios occurs only for higher
values of $i$, and thus it will be increasingly difficult to obtain precise
estimates for higher values of $\alpha$.

Even in cases where the singularity of physical interest is the one closest to
the origin, the d-log Pad\'e approximants usually lead to better estimates
than the ratio method \cite{g89}. The approximants are defined as ratios of
two polynomials $P_L(\lambda)$ and $Q_M(\lambda)$:
\begin{equation}
F_{LM}(\lambda)=\frac{P_L(\lambda)}{Q_M(\lambda)}=\frac{\sum_{i=0}^Lp_i
\lambda^i}{1+\sum_{j=1}^M q_j \lambda^j}.
\label{pa}
\end{equation}
The series for $\frac{d}{d \lambda} \ln P_\infty (\lambda)$ for fixed values of
$\alpha$ are substituted in the defining equation \ref{pa} and the
coefficients of the polynomials are chosen such that the identity is true up
to the order of the available series expansion. Thus approximants with $L+M
\leq 24$ may be built with the available series. Usually diagonal ($L=M$) or
close to diagonal approximants furnish better results, so we restricted our
calculations to these cases. The estimate for the critical value of $\lambda$
is found among the poles of the approximant, the estimate for the critical
exponent $\beta$ will be the residue at this pole.

We thus built approximants with $\alpha$ ranging between 0 and 40, estimating
the critical value of the parameter $\lambda$ as well as the critical exponent 
$\beta$. Although for small values of $\alpha$ the results are very good, with
estimates of $\beta$ comparable to the best ones in the literature for the
contact process, as the value of $p_c$ is decreased we notice a growing
dispersion of the estimates for the exponent and for $p_c<0.01$ most of the
approximants lead even to ill conditioned systems of linear equations for the
coefficients, and therefore we were not able to obtain estimates in this
region. We made additional one-variable investigations, such as obtaining
approximants for $(\lambda_c-\lambda)\frac{d}{d\lambda} P_s(\lambda)=\beta$
for several values of $\lambda_c$ and searching for the intercept of the curves
$\beta(\lambda_c)$ \cite{dj91} and non-homogeneous Pad\'e approximants
\cite{g89,nha}, and although some improvements of the estimates may be
obtained in 
certain cases, the situation does not change qualitatively. In figure
\ref{cline} the Pad\'e estimates for the critical line are displayed.

Actually, the increasing dispersion of the estimates as the parameter
$p_c$ approaches zero is not a surprise, since as was mentioned above in this
limit the model corresponds to the voter model which is in the compact
directed percolation (CDP) universality class, whose exponents are different
from the ones of the contact process, which belongs to the directed 
percolation (DP) universality class. For the voter model, the exponent $\beta$
of the order parameter is equal to zero (a discontinuity in the order parameter
occurs at the transition) \cite{h00}, but the exponent for the survival
probability $\beta^\prime$ is distinct from $\beta$ and equal to 1
\cite{dt95}. Therefore a crossover to from the DP to the CDP universality class
occurs as $p_c \to 0$, and it is known that in such situations the reduction
of two-variable series to one variable leads to very poor estimates
\cite{fk77}. So we analyzed the series without reducing the problem to one
variable, and to our knowledge the best results for two-variable series
applied to multicritical phenomena in the literature were obtained using the
partial differential approximants (PDA's) \cite{fk77}. They may be regarded as
a 
generalization to two variables of the d-log Pad\'e approximants. The defining
equation of the approximants is 
\begin{equation}
P_{\mathbf L}(x,y)F(x,y)=Q_{\mathbf M}(x,y)\frac{\partial F(x,y)}{\partial x} +
R_{\mathbf N}(x,y) \frac{\partial F(x,y)}{\partial y},
\label{pda}
\end{equation}
where $P$, $Q$, and $R$ are polynomials in the variables $x$ and $y$ with the
set of nonzero coefficients ${\mathbf L}$, ${\mathbf M}$, and ${\mathbf N}$,
respectively. The coefficients 
of the polynomials are obtained through substitution of the series expansion
for the quantity which is going to be analyzed 
\begin{equation}
f(x,y)=\sum_{k,k^\prime=0} f(k,k^\prime)x^k y^{k^\prime}
\end{equation}
into the defining equation \ref{pda} and requiring the equality to hold for a
set of indexes defined as ${\mathbf K}$. Again this procedure leads to a system
of linear 
equations for the coefficients of the polynomials, and since the coefficients
$f_{k,k^\prime}$ of the series are known for a finite set of indexes this sets
an upper limit to the number of coefficients in the polynomials. Since the
number of equations has to match the number of unknown coefficients, we must
have that the numbers of elements in each set satisfy $K=L+M+N-1$ (one
coefficient is fixed arbitrarily). An additional issue, 
which is not present in the one-variable case, is the symmetry of the
polynomials. Two frequently used options are the triangular and the
rectangular arrays of coefficients. The choice of these symmetries is related
to the symmetry of the series itself \cite{s90}. Below we discuss the solution
we adopted in the present case for this point.

Let us suppose that quantity represented by the series is expected to have a
multicritical behavior at a point $(x_c,y_c)$, described by
\begin{equation}
f(x,y) \approx |\Delta \widetilde{x}|^{-\gamma}Z\left( \frac{|\Delta
\widetilde{y}|}{|\Delta \widetilde{x}|^\phi} \right),
\label{mcs}
\end{equation}
where
\begin{equation}
\Delta \widetilde{x}=(x-x_c )-(y-y_c)/e_2,
\end{equation}
and
\begin{equation}
\Delta \widetilde{y}=(y-y_c)-e_1(x-x_c).
\end{equation}
Here $\gamma$ is the critical exponent of the quantity described by $f$ when
$\Delta \widetilde{y}=0$, $e_1$ and $e_2$ are the scaling slopes \cite{fk77}
and 
$\phi$ is the crossover exponent. The function $Z(z)$ is singular for one or
more values of its argument, corresponding to the critical line(s) incident on
the multicritical point. Once the coefficients of the defining polynomials are
obtained, the estimated location of the multicritical point corresponds to the
common zero of the polynomials $Q_M$ and $R_N$. This may be seen substituting
the scaling form \ref{mcs} in the defining equation \ref{pda} of the
approximant. The exponents and scaling
slopes may also be obtained directly from the polynomials, without
integrating the partial differential equation. A detailed discussion of the
algorithm, as well as computer codes, may be found in \cite{s90}.

Before proceeding with the analysis of the series, it is convenient to perform
a change of variables, since the multicritical point in the original variables
is located at $\alpha \to \infty$. We thus express the series in the variables 
\begin{eqnarray*}
x =  \lambda  =  \frac{2p_c}{p_a}\\
y =  \alpha\lambda  =  \frac{1-p_a-p_c}{p_a}.
\end{eqnarray*}
In these new variables the multicritical point is located at $x=0$, $y=1$, and
the survival probability may be written as
\begin{equation}
P_{\infty}=1-\frac{1}{2}x-y-xF(x,y),
\label{ns}
\end{equation}
where
\begin{equation}
F(x,y)=\sum_{i=1}^{24}\sum_{j=0}^i b_{i+1,j}x^{i-j}y^j
\end{equation}
is represented by a series with triangular symmetry, which is very convenient
to be analyzed using PDA's. The number of approximants which may be obtained
from the series is very big, so we restricted ourselves to approximants with
the number of elements in $\mathbf{M}$ close to the number of elements in
$\mathbf{N}$. The 
polynomials had the same triangular symmetry as the series, but in most cases
some higher order elements of the polynomials were set equal to zero in order
to 
match the number of unknowns to the number of equations in the set of linear
equations for the coefficients. This is a rather standard procedure and is
discussed in detail by Styer \cite{s90}. Even at
rather low orders, we found a reasonable agreement between most of the
estimates from different approximants. Finally,
we considered a set of 42 approximants which use the elements of the series
for $F(x,y)$ with highest orders $i$ between 15 and 24. Discarding some
approximants which generated estimates which were rather far away of the
general trend, finally we used a set of 36 approximants to obtain the
estimates.

\begin{center}
\begin{figure}
\includegraphics[scale=0.7]{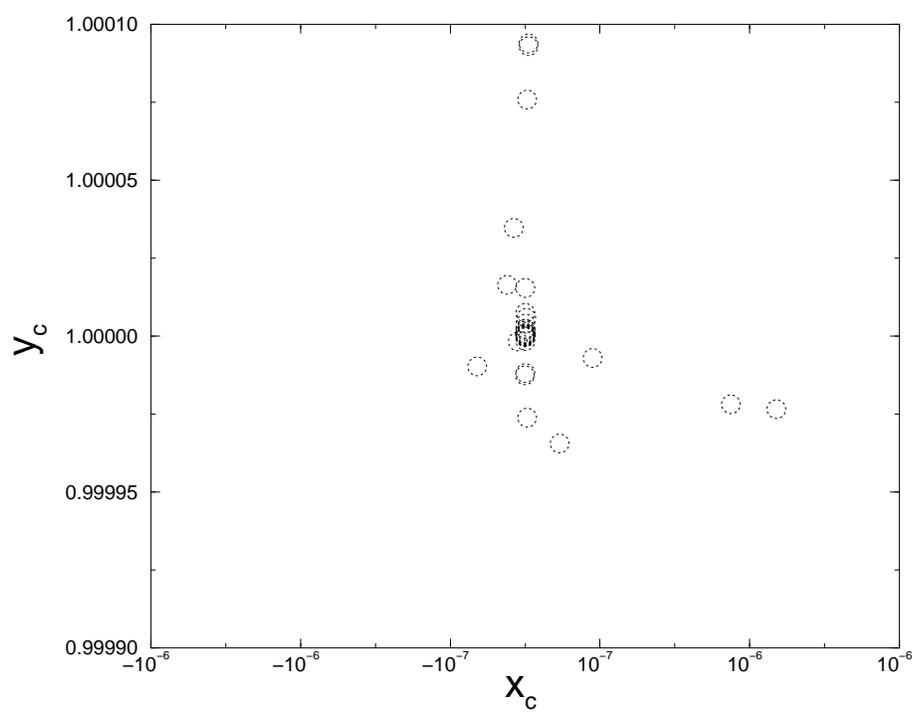}
\caption{Estimates for the location of the multicritical point provided by the
set of PDA's.}
\label{mcp}
\end{figure}
\end{center}

\begin{center}
\begin{figure}
\includegraphics[scale=0.7]{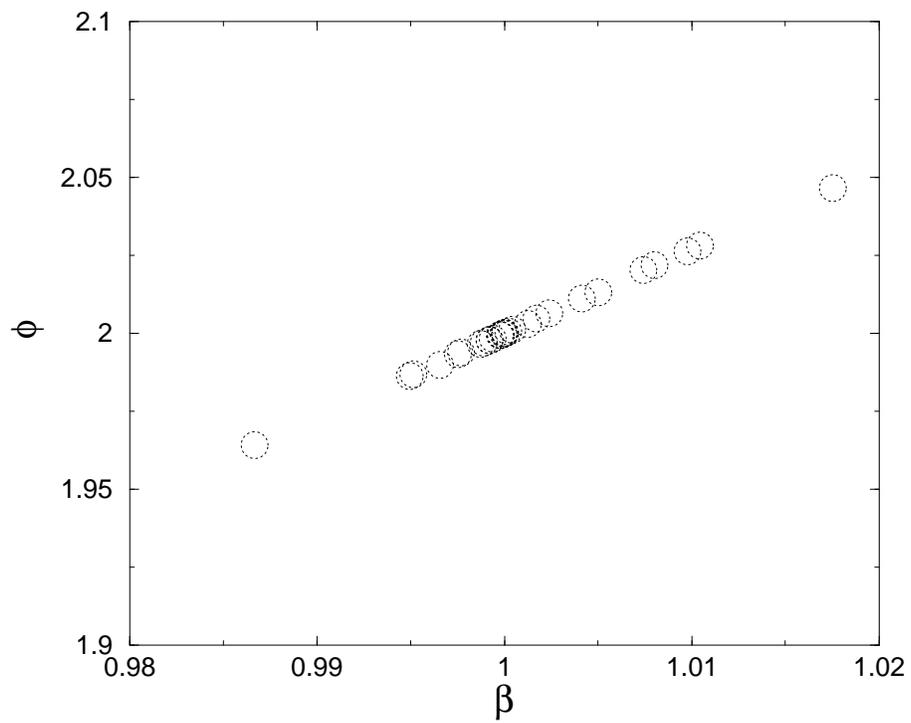}
\caption{Estimates for the values of the exponents $\beta^\prime$ and $\phi$.}
\label{betaphi}
\end{figure}
\end{center}

In figure \ref{mcp} the estimates for the location of the multicritical point
are displayed. We notice that the estimates are very close to the exactly known
values $x=0$ and $y=1$. The estimated values for $x_c$ and $y_c$ are $(0.4 \pm
1.8) \times 10^{-6}$ and $y=1.0000 \pm 0.0003$. The exponent $-\gamma$ in the
multicritical scaling form \ref{mcs} corresponds to the exponent
$\beta^\prime$ of the CDP universality class. The estimated value is equal to
$1.00 \pm 0.01$, which agrees very well with the exact value $\beta^\prime=1$
\cite{h00}. Finally, the crossover exponent was estimated as $\phi=2.01 \pm
0.03$, and thus the mean field value for this exponent ($\phi=2$ 
\cite{dts03})
is within the confidence interval of our estimate. The estimates for
$\beta^\prime$ and $\phi$ are shown in figure \ref{betaphi}. We also obtained
biased PDA's, fixing the other parameters at their known values and calculating
improved estimates for $\phi$. This procedure resulted in the estimate
$\phi=2.00 \pm 0.02$. We thus conclude that our estimates for $\phi$ are very
close to the classical value of the crossover exponent. The estimates for the
slopes of the scaling axes show a rather broad distribution, a significant
majority 
of the approximants provide quite large values for $e_1$, while $e_2$ is
typically much smaller. This suggests that $\Delta \widetilde{y}=x$ and it is
reasonable to choose $\Delta \widetilde{x}=1-y$, since it corresponds to the
weak direction, parallel to the critical line at the multicritical point. In
the limit of the voter model ($x=0$) we notice that the series \ref{ns}
reduces to the exact result $P_\infty=1-y$.

One way to actually estimate values of the quantity which is described by a
PDA is to integrate the equation using the method of characteristics. A
timelike variable $\tau$ is defined and a family of curves $(x(\tau),y(\tau))$
(the characteristics) is considered. These curves are defined by the
equations
\begin{eqnarray}
\frac{dx}{d \tau} &=& Q_M(x(\tau),y(\tau)),\nonumber\\
\frac{dy}{d\tau}  &=& R_N(x(\tau),y(\tau)).
\label{char}
\end{eqnarray}
Along such a curve, the defining equation of the PDA \ref{pda} leads to an
ordinary differential equation for $F$.
\begin{equation}
\frac{dF}{d\tau}=P_L(x(\tau),y(\tau))F,
\end{equation}
which may readily be integrated, providing the value of $F$ at the points of
the characteristics, once we know this value at a initial point. Our efforts
to obtain estimates for the survival probability, particularly close to the
multicritical point, using this procedure were not very
successful. Nevertheless, it is worth to mention that an estimate for the
critical line may be obtained this way. The critical line, which connects the
point which corresponds to the CP to the multicritical point of the voter
model transition is a line of singularities, and it is not difficult to show
that such a line is one of the characteristics defined by equations \ref{char}.
Therefore, the characteristic whose initial point is located on the transition
point of the CP, which is known with great accuracy, is an estimate for the
critical line. We therefore chose a particular approximant with estimates
close to the mean values. The number of elements in the sets for this
approximant are $K=300$, $L=171$, and $M=N=65$. In figure \ref{cline} this
characteristic curve is shown, and a 
nice agreement with other estimates of the critical line may be observed.

\section{Conclusion and discussion}
\label{conclu}
\begin{center}
\begin{figure}
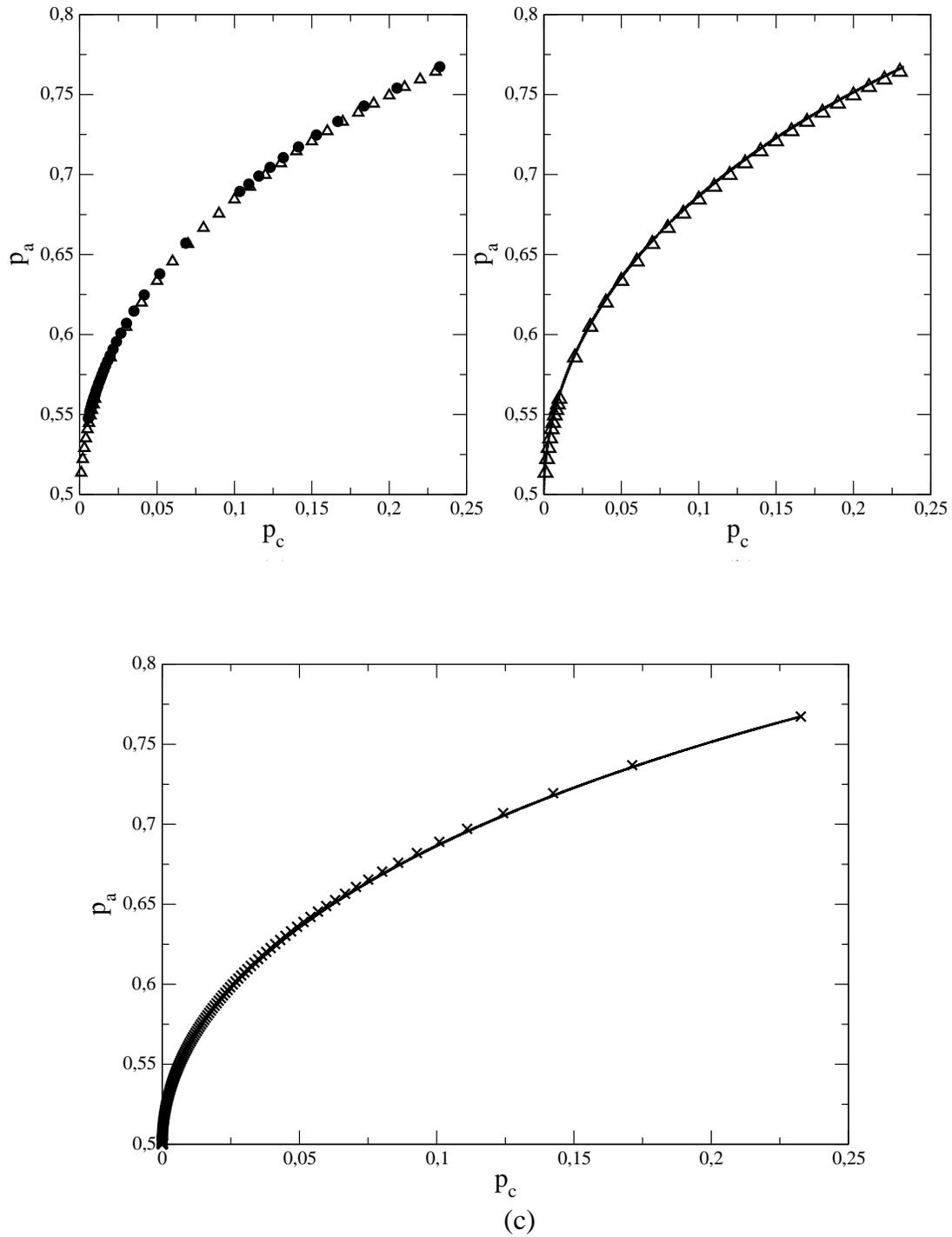

\includegraphics[scale=0.5]{compsim.eps}

\vspace{1cm}

\includegraphics[scale=0.5]{compcarac.eps}
\caption{Estimates for the critical line. Curve (a) shows results from
simulations (triangles) and Pad\'e approximants (circles). Curve (b) displays
the characteristic which starts at the CP (full line) and simulation results
(triangles). In curve (c) the full line is again the characteristic and the
values which follow from the multicritical scaling form are represented by
crosses.}
\label{cline}
\end{figure}
\end{center}

We may notice that
the coefficients $b(i,i-1)$, for $i=1,2,\ldots$ are equal to $1/2$, a 
result
which is indicated by the coefficients below but may be shown to be true in
general using the operator formalism above. Therefore, we may sum these set 
of terms in the series, obtaining
\begin{equation}
P_\infty=(1-y) \left[1-\frac{1}{2}\frac{x}{(1-y)^2}- \ldots \right].
\end{equation}
Now if we compare this expression with the multicritical scaling form 
\ref{mcs}, we may recognize between braces the two first terms of a 
Taylor expansion of the multicritical scaling function $Z(z)$, where 
the variable 
$z$ is identified as $\frac{x}{(1-y)^2}$. This agrees with the estimates 
obtained 
from the PDA's. Moreover, remembering that for $y=0$ the series 
represents the CP,
the function $Z(z)$ may be recognized as the survival probability of the CP,
which was studied in great detail \cite{dj91} and found to have a singularity
at $z_0=0.6064$ with the exponent $\beta^\prime=\beta=0.276486$. As expected, 
the 
critical line is characterized by the same exponent of the CP, that is, is
in the DP universality class. The critical line corresponds to $z=z_0$, that is
\begin{equation}
x=z_0(1-y)^2,
\end{equation}
and this curve is shown in figure \ref{cline}. It is interesting to notice
that  
the agreement of this estimate of the critical line with the other two,
obtained from  
Pad\'e and partial differential approximants, is quite good even far away from
the multicritical point. The multicritical scaling form with the
identification of the scaling variable $z$ above is exact in the limit $p_c=0$
(biased voter model) and reproduces the supercritical series expansion for the
CP, when $y=0$. It does, however, not reproduce the two-variable supercritical
series expansion for the full model we considered here away from these
limiting cases. 

The Domany Kinzel probabilistic cellular automaton
\cite{dk84} in part of its two-dimensional phase diagram corresponds to a
synchronous update version of the contact process, and as stated above has the
synchronous update voter model as a endpoint of the critical line. It is
believed, although to our knowledge not proven, that if in a particular model
the update procedure is changed from synchronous to asynchronous, the
critical exponents do not change.Some bounds for the critical line are
presented in \cite{kt95}. Although an upper bound for this line due to Liggett
\cite{l94} is quadratic close to the CDP endpoint, the lower bounds are
linear, and thus the asymptotic behavior of the critical line is not fixed by
those bounds. The critical line is studied in more detail by simulations and
series expansions in \cite{tikt95}, and based on this results the authors
conjectured a quadratic asymptotic behavior of the critical line, consistent
with $\phi=2$.Finally, a more detailed series analysis is done in
\cite{ikb01}, but since one-variable Pad\'e approximants were used, the
results are not precise in the region close to the CDP point. Thus, there are
indications that the crossover exponent has the same value in both models,
and if these indications are correct, the invariance of this multicritical
exponent with respect to the update procedure is verified in this particular
case.

In conclusion, the analysis of the series for the ultimate survival
probability using PDA's lead to quite precise
estimates for the multicritical exponents, and these estimates, together with
the possibility to sum the terms of the two-variable series which are linear
in $x$ allowed us to conjecture the exact form of the multicritical scaling
expression. The multicritical scaling function $Z(z)$ is known as a series
expansion up to order 25.

\begin{acknowledgments}
We thank Prof. Ronald Dickman for many helpful discussions, and one of the
referees for calling out attention to references \cite{kt95}
-\cite{ikb01}. We
also thank Professor M\'ario J. de Oliveira for a critical reading of the
manuscript. This 
research was partially supported by the Brazilian agencies CAPES, FAPERJ
and CNPq, particularly through the project PRONEX-CNPq-FAPERJ/171.168-2003.
\end{acknowledgments}

\begin{center}
\begin{table}
\begin{tabular}{|c|c|c|}
\hline
$i$&$j$&$b_{i,j}$\\
\hline
1&1&    0.10000000000000000000D+01\\ 
1&0&    0.50000000000000000000D+00\\ 
\hline
2&1&    0.50000000000000000000D+00\\
2&0&    0.25000000000000000000D+00\\
\hline
3&2&    0.50000000000000000000D+00\\
3&1&    0.75000000000000000000D+00\\
3&0&    0.25000000000000000000D+00\\
\hline
4&3&    0.50000000000000000000D+00\\
4&2&    0.15000000000000000000D+01\\
4&1&    0.11875000000000000000D+01\\
4&0&    0.28125000000000000000D+00\\
\hline
5&4&    0.50000000000000000000D+00\\
5&3&    0.25000000000000000000D+01\\
5&2&    0.35234375000000000000D+01\\
5&1&    0.18867187500000000000D+01\\
5&0&    0.34375000000000000000D+00\\
\hline
6&5&    0.50000000000000000000D+00\\
6&4&    0.37500000000000000000D+01\\
6&3&    0.81665039062499928945D+01\\
6&2&    0.73991699218749946709D+01\\
6&1&    0.29899902343749985789D+01\\
6&0&    0.44726562499999982236D+00\\
\hline
7&6&    0.50000000000000000000D+00\\
7&5&    0.52499999999999982236D+01\\
7&4&    0.16280639648437500000D+02\\
7&3&    0.21964294433593751776D+02\\
7&2&    0.14622192382812501776D+02\\
7&1&    0.47470703125000008881D+01\\
7&0&    0.60223388671874964472D+00\\
\hline
8&7&    0.50000000000000000000D+00\\
8&6&    0.69999999999999982236D+01\\
8&5&    0.29241512298583977269D+02\\
8&4&    0.54546823501586914062D+02\\
8&3&    0.52912919998168961299D+02\\
8&2&    0.27873636245727544391D+02\\
8&1&    0.75799846649169939638D+01\\
8&0&    0.83485031127929723027D+00\\
\hline
9&8&    0.50000000000000000000D+00\\
9&7&    0.90000000000000017763D+01\\
9&6&    0.48670531988143883594D+02\\
9&5&    0.11948751342296599631D+03\\
9&4&    0.15729669463192959000D+03\\
9&3&    0.11890031438403647712D+03\\
9&2&    0.51820946525644346891D+02\\
9&1&    0.12126680864228140954D+02\\
9&0&    0.11814667913648822050D+01\\
\hline
10&9&    0.50000000000000000000D+00\\
10&8&    0.11249999999999982236D+02\\
10&7&    0.76407402418553784784D+02\\
10&6&    0.23825064151361545761D+03\\
10&5&    0.40643453331768384373D+03\\
10&4&    0.41173626079900831342D+03\\
10&3&    0.25488795003760951196D+03\\
10&2&    0.94770005067848632762D+02\\
10&1&    0.19463028091937289332D+02\\
10&0&    0.16988672076919923981D+01\\
\hline
\end{tabular}
\end{table}
\end{center}
\begin{center}
\begin{table}
\begin{tabular}{|c|c|c|}
\hline
11&10&    0.50000000000000000000D+00\\
11&9&    0.13750000000000004440D+02\\
11&8&    0.11453289534710346941D+03\\
11&7&    0.44150459107663433400D+03\\
11&6&    0.94395713065082347270D+03\\
11&5&    0.12251638531684152511D+04\\
11&4&    0.10067004627495166335D+04\\
11&3&    0.52744546391776081506D+03\\
11&2&    0.17102490738153186100D+03\\
11&1&    0.31313301122887615690D+02\\
11&0&    0.24775957118665874467D+01\\
\hline
12&11&    0.50000000000000000000D+00\\
12&10&    0.16500000000000039079D+02\\
12&9&    0.16534952380310286024D+03\\
12&8&    0.77136775289458459070D+03\\
12&7&    0.20158736418235778664D+04\\
12&6&    0.32476491678046719435D+04\\
12&5&    0.33907974883298468427D+04\\
12&4&    0.23413097838386356386D+04\\
12&3&    0.10630662385637921207D+04\\
12&2&    0.30555868520944353683D+03\\
12&1&    0.50452588505226367843D+02\\
12&0&    0.36488812342264482779D+01\\
\hline
13&12&    0.50000000000000000000D+00\\
13&11&    0.19499999999999982236D+02\\
13&10&    0.23139824792517162954D+03\\
13&9&    0.12840079302994218402D+04\\
13&8&    0.40209683861219263079D+04\\
13&7&    0.78496665288886928735D+04\\
13&6&    0.10096110998534653102D+05\\
13&5&    0.88003157466952828258D+04\\
13&4&    0.52368347467192908339D+04\\
13&3&    0.20980007339714172864D+04\\
13&2&    0.54172526170510240106D+03\\
13&1&    0.81489527352144666139D+02\\
13&0&    0.54293656084851633636D+01\\
\hline
14&13&    0.49999986588954925537D+00\\
14&12&    0.22749997384846243342D+00\\
14&11&    0.31544379761375296311D+03\\
14&10&    0.20523899291719041038D+04\\
14&9&    0.75794568674053097723D+04\\
14&8&    0.17597122905598574504D+05\\
14&7&    0.27244155488205645809D+05\\
14&6&    0.29082737294330742727D+05\\
14&5&    0.21732772717178829857D+05\\
14&4&    0.11357034171626549934D+05\\
14&3&    0.40697753516544512564D+04\\
14&2&    0.95363664836240946698D+03\\
14&1&    0.13169793741584487900D+03\\
14&0&    0.81325422193071297272D+01\\
\hline
\end{tabular}
\end{table}
\end{center}
\begin{center}
\begin{table}
\begin{tabular}{|c|c|c|}
\hline
15&14&    0.50000000000000000000D+00\\
15&13&    0.26249998256564137655D+02\\
15&12&    0.42048775894871077696D+03\\
15&11&    0.31693914938186833474D+04\\
15&10&    0.13620079986016961903D+05\\
15&9&    0.37037632746719575393D+05\\
15&8&    0.67783286393305548500D+05\\
15&7&    0.86632679492839894663D+05\\
15&6&    0.78910368357158162666D+05\\
15&5&    0.51571902934753062197D+05\\
15&4&    0.24020987390678723016D+05\\
15&3&    0.77874939092949517771D+04\\
15&2&    0.16707029021016392533D+04\\
15&1&    0.21330513440327356633D+03\\
15&0&    0.12275012836144905126D+02\\
\hline
16&15&    0.50000000000000000000D+00\\
16&14&    0.29999999329447643248D+02\\
16&13&    0.54975682655068318638D+03\\
16&12&    0.47510546519798584341D+04\\
16&11&    0.23493882045888385690D+05\\
16&10&    0.73897182789227748855D+05\\
16&9&    0.15755031904987298219D+06\\
16&8&    0.23687090513090347521D+06\\
16&7&    0.25718644037330271601D+06\\
16&6&    0.20406058122209760341D+06\\
16&5&    0.11843376024372214061D+06\\
16&4&    0.49739262723985895320D+05\\
16&3&    0.14718788444797093362D+05\\
16&2&    0.29113789219292591781D+04\\
16&1&    0.34558659823607458250D+03\\
16&0&    0.18620961415130533822D+02\\
\hline
17&16&    0.50000000000000000000D+00\\
17&15&    0.33999999329447789797D+02\\
17&14&    0.70671335354464535072D+03\\
17&13&    0.69402158602317101099D+04\\
17&12&    0.39112396145099372901D+05\\
17&11&    0.14079076808597497105D+06\\
17&10&    0.34548310262719810204D+06\\
17&9&    0.60241714803092252239D+06\\
17&8&    0.76636402935447742734D+06\\
17&7&    0.72224826262886052674D+06\\
17&6&    0.50729839736110484693D+06\\
17&5&    0.26473279696452087783D+06\\
17&4&    0.10122975226631107936D+06\\
17&3&    0.27555943803143052583D+05\\
17&2&    0.50567133817482794455D+04\\
17&1&    0.56084260997641584012D+03\\
17&0&    0.28405733950687368505D+02\\
\hline
\end{tabular}
\end{table}
\end{center}
\begin{center}
\begin{table}
\begin{tabular}{|c|c|c|}
\hline
18&17&    0.50000000000000000000D+00\\
18&16&    0.38249999329447760487D+02\\
18&15&    0.89504578674400576687D+03\\
18&14&    0.99102626626136949283D+04\\
18&13&    0.63122994123535232091D+05\\
18&12&    0.25768737165396164989D+06\\
18&11&    0.72043302015833452500D+06\\
18&10&    0.14399790608479312581D+07\\
18&9&    0.21166673049189457245D+07\\
18&8&    0.23292641415436174945D+07\\
18&7&    0.19371723268429082764D+07\\
18&6&    0.12198708033512473125D+07\\
18&5&    0.57806709777733367161D+06\\
18&4&    0.20284284513882400169D+06\\
18&3&    0.51115204972041548003D+05\\
18&2&    0.87477997069442139377D+04\\
18&1&    0.91057218719788650673D+03\\
18&0&    0.43520782874229171355D+02\\
\hline
19&18&    0.49999999999999911182D+00\\
19&17&    0.42749999329447687657D+02\\
19&16&    0.11186774875974003773D+04\\
19&15&    0.13869278218084735243D+05\\
19&14&    0.99114912109387987015D+05\\
19&13&    0.45525829782828974856D+06\\
19&12&    0.14375483791065075678D+07\\
19&11&    0.32614349098596422393D+07\\
19&10&    0.54767322733604206774D+07\\
19&9&    0.69423928844342714938D+07\\
19&8&    0.67230154917519895363D+07\\
19&7&    0.50002124204980198385D+07\\
19&6&    0.28527370161475179344D+07\\
19&5&    0.12379379646484691690D+07\\
19&4&    0.40134571279240418562D+06\\
19&3&    0.94145168822797273833D+05\\
19&2&    0.15093606456186623887D+05\\
19&1&    0.14798263077956907984D+04\\
19&0&    0.66930218067927942371D+02\\
\hline
20&19&    0.50000000000000000000D+00\\
20&18&    0.47499999329447701867D+02\\
20&17&    0.13817585309510260671D+04\\
20&16&    0.19064302620655780629D+05\\
20&15&    0.15187519641466302289D+06\\
20&14&    0.77950921221482092349D+06\\
20&13&    0.27591465361185987248D+07\\
20&12&    0.70457626117426288558D+07\\
20&11&    0.13386942665323722234D+08\\
20&10&    0.19328829379718762027D+08\\
20&9&    0.21502873943167273296D+08\\
20&8&    0.18574978781954090578D+08\\
20&7&    0.12487211289613961984D+08\\
20&6&    0.65099543472194438820D+07\\
20&5&    0.26049731385271019945D+07\\
20&4&    0.78478644916613982118D+06\\
20&3&    0.17219353528307646428D+06\\
20&2&    0.25969733685492251140D+05\\
20&1&    0.24071073326750083154D+04\\
20&0&    0.10337399908888862398D+03\\
\hline
\end{tabular}
\end{table}
\end{center}
\begin{center}
\begin{table}
\begin{tabular}{|c|c|c|}
\hline
21&20&    0.49999999999999955591D+00\\
21&19&    0.52499999329447701867D+02\\
21&18&    0.16886733349218010502D+04\\
21&17&    0.25785991340403939808D+05   \\
21&16&    0.22768166104993685650D+06   \\
21&15&    0.12978288769859755902D+07   \\
21&14&    0.51154797613905129693D+07   \\
21&13&    0.14596468511940512868D+08   \\
21&12&    0.31124330624461475913D+08   \\
21&11&    0.50710951504426553526D+08   \\
21&10&    0.64100799162967723177D+08   \\
21&9&    0.63467922470529245515D+08   \\
21&8&    0.49453964458225394551D+08   \\
21&7&    0.30321310892616786247D+08   \\
21&6&    0.14550387399195248150D+08   \\
21&5&    0.54012506504483805969D+07   \\
21&4&    0.15195548211633345125D+07   \\
21&3&    0.31313135707735888502D+06   \\
21&2&    0.44571407092920996007D+05   \\
21&1&    0.39157750249797601327D+04   \\
21&0&    0.15998830271630990473D+03   \\
\hline
22&21&    0.50000000000000088817D+00   \\
22&20&    0.57749999329447438967D+02   \\
22&19&    0.20440335494663091076D+04   \\
22&18&    0.34373382551268485407D+05   \\
22&17&    0.33466104195644446051D+06   \\
22&16&    0.21070967788898307126D+07   \\
22&15&    0.91945254038400960894D+07   \\
22&14&    0.29128975460417066756D+08   \\
22&13&    0.69213935104954478205D+08   \\
22&12&    0.12623609387336516274D+09   \\
22&11&    0.17963375603295976823D+09   \\
22&10&    0.20164422581600076611D+09   \\
22&9&    0.17971976143434213568D+09   \\
22&8&    0.12747436997852454876D+09   \\
22&7&    0.71815965145384188517D+08   \\
22&6&    0.31918963164481914951D+08   \\
22&5&    0.11049035068083481458D+08   \\
22&4&    0.29157961094487756525D+07   \\
22&3&    0.56650954312689849601D+06   \\
22&2&    0.76372602458216514165D+05   \\
22&1&    0.63802269924577696968D+04   \\
22&0&    0.24876519640924477094D+03   \\
\hline
\end{tabular}
\end{table}
\end{center}
\begin{center}
\begin{table}
\begin{tabular}{|c|c|c|}
\hline
23&22&    0.50000015522128578027D+00   \\
23&21&    0.63250003986086227314D+02   \\
23&20&    0.24526847667386335594D+04   \\
23&19&    0.45219068649736344767D+05   \\
23&18&    0.48319032233047343183D+06   \\
23&17&    0.33439431430492136954D+07   \\
23&16&    0.16069927289546637183D+08   \\
23&15&    0.56208076057486060506D+08   \\
23&14&    0.14791153728125157051D+09   \\
23&13&    0.29990967758279261090D+09   \\
23&12&    0.47669571268868358160D+09   \\
23&11&    0.60118669013109657939D+09   \\
23&10&    0.60634053785517059154D+09   \\
23&9&    0.49107989528314268667D+09    \\
23&8&    0.31953776898326688993D+09    \\
23&7&    0.16647521279074794620D+09    \\
23&6&    0.68894669993720274447D+08    \\
23&5&    0.22337519908203442575D+08    \\
23&4&    0.55496488699702037905D+07    \\
23&3&    0.10196528668928661609D+07    \\
23&2&    0.13050250676997474652D+06    \\
23&1&    0.10386065610378922841D+05    \\
23&0&    0.38696067609363534955D+03    \\
\hline
24&23&    0.49999999999999706901D+00   \\
24&22&    0.69000000457884933524D+02   \\
24&21&    0.29197000139992891121D+04   \\
24&20&    0.58774468615505739421D+05   \\
24&19&    0.68638127471201864082D+06   \\
24&18&    0.51981592602471167197D+07   \\
24&17&    0.27383018956536124832D+08   \\
24&16&    0.10521292804690998146D+09   \\
24&15&    0.30494730611982263646D+09   \\
24&14&    0.68324606257180731105D+09   \\
24&13&    0.12048047382312678799D+10   \\
24&12&    0.16938688975594596186D+10   \\
24&11&    0.19158522496513690214D+10   \\
24&10&    0.17528882189096098187D+10   \\
24&9&    0.13003099701647686803D+10    \\
24&8&    0.78129739240902704722D+09    \\
24&7&    0.37851359502279247060D+09    \\
24&6&    0.14654979817717892487D+09    \\
24&5&    0.44688725216130675832D+08    \\
24&4&    0.10490868838890698988D+08    \\
24&3&    0.18288174674782467832D+07    \\
24&2&    0.22289668577994086184D+06    \\
24&1&    0.16948463518651144532D+05    \\
24&0&    0.60509199550607846163D+03    \\
\hline
\end{tabular}
\end{table}
\end{center}
\begin{center}
\begin{table}
\begin{tabular}{|c|c|c|}
\hline
25&24&    0.50000051317104317050D+00   \\
25&23&    0.75000019616897146690D+02   \\
25&22&    0.34503863852011384949D+04   \\
25&21&    0.75555516468201169288D+05   \\
25&20&    0.96061396122189321999D+06   \\
25&19&    0.79292390146647733217D+07   \\
25&18&    0.45593449242937929000D+08   \\
25&17&    0.19157385641492643557D+09   \\
25&16&    0.60860240979789494986D+09   \\
25&15&    0.14987944522389162749D+10   \\
25&14&    0.29148047590757961700D+10   \\
25&13&    0.45381694251898121450D+10   \\
25&12&    0.57125179306761664221D+10   \\
25&11&    0.58520654724976619576D+10   \\
25&10&    0.48967910880832077324D+10   \\
25&9&    0.33495648941633358042D+10    \\
25&8&    0.18689764057712466183D+10    \\
25&7&    0.84597759965914463009D+09    \\
25&6&    0.30764609873623349756D+09    \\
25&5&    0.88524517163033653588D+08    \\
25&4&    0.19689926825686265843D+08    \\
25&3&    0.32638605561858318182D+07    \\
25&2&    0.37947216762104512000D+06    \\
25&1&    0.27603268426234146559D+05    \\
25&0&    0.94519638900755200694D+03    \\
\hline
\end{tabular}
\caption{Coefficients $b_{i,j}$ of the supercritical series expansion for the
ultimate survival probability of A particles.}
\label{coefs}
\end{table}
\end{center}

\end{document}